\input harvmac
\overfullrule=0pt

%

\def\bar{\overline}

\def\K3{{\bf K3}}

\Title{ \vbox{\baselineskip12pt
\hbox{hep-th/0007212}
\hbox{IFP-0007-UNC}
\hbox{HUB-EP-00/18}
\hbox{CPHT-RR 006.0300}}}
{\vbox{\centerline{Anomalous D-Brane Charge in F-Theory Compactifications}}}
\centerline{Bj\"orn Andreas$^{\dagger}$\foot{bandreas@physics.unc.edu, 
supported by U.S. DOE grant DE-FG05-85ER40219}, Gottfried Curio$^*$\foot{
curio@physik.hu-berlin.de} and 
Ruben Minasian$^{\ddagger}$\foot{ruben@cpht.polytechnique.fr}}
\smallskip
\centerline{\it $^{\dagger}$Department of Physics and Astronomy}
\centerline{\it
University of North Carolina, Chapel Hill, NC 27599-3255, USA}
\smallskip
\centerline{\it $^*$Humboldt-Universit\"at zu Berlin,}
\centerline{\it Institut f\"ur Physik, D-10115 Berlin, Germany}
\smallskip
\centerline{\it $^{\ddagger}$Centre de Physique Th{\'e}orique,
Ecole Polytechnique,  F-91128 Palaiseau, France}
\centerline{\it
Unit\'e mixte du CNRS et de l'EP, UMR 7644}
\bigskip
\def\sqr#1#2{{\vbox{\hrule height.#2pt\hbox{\vrule width
.#2pt height#1pt \kern#1pt\vrule width.#2pt}\hrule height.#2pt}}}

\noindent Tadpole cancellation in F-theory on an
elliptic Calabi-Yau fourfold $X\rightarrow B_3$ demands some
spacetime-filling three-branes (points in $B_3$). If moved to the 
discriminant surface, which supports the gauge group,
and dissolved into a finite size instanton, the second Chern class of
the corresponding bundle $E$ is expected to give 
a compensating contribution. However the dependence of D-brane
charge on the geometry of $W$ and on the embedding $i: W\rightarrow
B_3$ gives a correction to $c_2(E)$. We show how this is reconciled by
considering the torsion sheaf $i_*E$ and  discuss some integrality issues
related to global properties of $X$ as well as the moduli space of this 
object.

\Date{July, 2000}

\lref\Fvafa{C. Vafa, ``Evidence for F-Theory'', Nucl.Phys. {\bf B469} (1996)
403, hep-th/9602022.}

\lref\MV{D.R. Morrison and C. Vafa, ``Compactifications of F-Theory on 
Calabi--Yau Threefolds -- I'', Nucl.Phys. {\bf B473} (1996) 74,
hep-th/9602114; ``Compactifications of F-Theory on 
Calabi--Yau Threefolds -- II'', Nucl.Phys. {\bf B476} (1996) 437, 
hep-th/9603161.}

\lref\FMW{R. Friedman, J. Morgan and E. Witten, ``Vector Bundles and F- 
Theory'', Commun. Math. Phys. {\bf 187} (1997) 679, hep-th/9701162.}

\lref\wflux{E. Witten, ``On Flux Quantization in M-Theory and the Effective
Action'', J. Geom. Phys. {\bf 22} (1997) 1, hep-th/9609122.}

\lref\damu{K. Dasgupta and S. Mukhi, ``A Note on Low Dimensional String
Compactifications'', Phys. Lett. {\bf B398} (1997) 285, hep-th/9612188.}

\lref\SVW{S. Sethi, C. Vafa and E. Witten, ``Constraints on Low Dimensional 
String Compactifications'', Nucl. Phys. {\bf B480} (1996) 213, 
hep-th/9606122.}

\lref\bersh{M. Bershadsky, A. Johansen, T. Pantev, V. Sadov, 
``On Four-Dimensional Compactifications of F-Theory'', 
Nucl.Phys. {\bf B505} (1997) 165-201, hep-th/9701165.}

\lref\mm{ R. Minasian and G. Moore,``K-theory and Ramond-Ramond charge'',
{\bf JHEP 9711} (1997) 002, hep-th/9710230}

\lref\dlm{M.~J.~Duff, J.~T.~Liu and R.~Minasian,
``Eleven-dimensional origin of string / string duality: A one-loop test,''
Nucl.\ Phys.\  {\bf B452}, 261 (1995), hep-th/9506126.}

\lref\drs{K. Dasgupta, G. Rajesh and S. Sethi, 
``M Theory, Orientifolds and G-Flux'', JHEP 9908 (1999) 023, hep-th/9908088.}

\lref\vw{C.~Vafa and E.~Witten,
``A One loop test of string duality,''
Nucl.\ Phys.\  {\bf B447}, 261 (1995), hep-th/9505053.}

\lref\fms{S.~Ferrara, R.~Minasian and A.~Sagnotti,
``Low-Energy Analysis of $M$ and $F$ Theories on Calabi-Yau
Threefolds'',
Nucl.\ Phys.\  {\bf B474}, 323 (1996), hep-th/9604097.}

\lref\ghm{M.~B.~Green, J.~A.~Harvey and G.~Moore,
``I-brane inflow and anomalous couplings on D-branes'',
Class.\ Quant.\ Grav.\  {\bf 14}, 47 (1997), hep-th/9605033.}

\lref\chy{Y.~E.~Cheung and Z.~Yin,
``Anomalies, branes, and currents'',
Nucl.\ Phys.\  {\bf B517}, 69 (1998), hep-th/9710206.}

\lref\szabo{K.~Olsen and R.~J.~Szabo, 
``Constructing D-Branes from K-Theory'', hep-th/9907140.}

\lref\bb{K.~Becker and M.~Becker,
``M-Theory on Eight-Manifolds'',
Nucl.\ Phys.\  {\bf B477}, 155 (1996), hep-th/9605053.}

\lref\AC{B. Andreas and G. Curio,
``On Discrete Twist and Four-Flux in N = 1 Heterotic/F-theory
 Compactifications'', hep-th/9908193.}

\lref\GVW{S. Gukov, C. Vafa and E. Witten ``CFT's From Calabi-Yau Four Folds'',
hep-th/9906070.}

\lref\hart{R. Hartshorne, ``Algebraic Geometry'', Springer 1977.}

\lref\Karo{F. Hirzebruch, ``Topological Methods in Algebraic Geometry'',
Springer 1978\semi
M. Karoubi, ``K-Theory'', Springer 1978.}

\lref\wik{E. Witten, ``D-Branes and K-theory, hep-th/9810188.}

\lref\koba{S. Kobayashi, ``Differential Geometry of Complex Vector Bundles'',
Princeton University Press 1987.}

\lref\donal{S. K. Donaldson, ``A polynomial Invariant for Smooth 
Four-Manifolds'', Topology {\bf 29} (1990) 257.}

\lref\thom{R. P. Thomas, ``A Holomorphic Casson Invariant For Calabi-Yau
Three-Folds, and Bundles on K3 Fibrations'', math.AG/9806111.}

\lref\hm{J.~A.~Harvey and G.~Moore, ``On the algebras of BPS states,''
Commun.\ Math.\ Phys.\  {\bf 197}, 489 (1998), hep-th/9609017.}

\lref\fw{D.~S.~Freed and E.~Witten,
``Anomalies in string theory with D-branes,''
hep-th/9907189.}  
%


\def\x{{\times}}


Two well-studied models for $N=1$ string compactifications are the heterotic
string on a Calabi-Yau three-fold with a vector bundle embedded in 
$E_8\x E_8$ and $F$-theory on a Calabi-Yau four-fold $X$ (in the
following when referring to cohomological formulas involving $X$ or
couplings given by integration over $X$, we denote by $X$ the 
smooth resolved four-fold) \refs{\Fvafa, \MV, \bersh} . In the last case 
$X$ is an elliptic fibration over $B_3$ which in turn is a ${\bf P}^1$ 
fibration\foot{The fibration structure of $\pi: B_3\rightarrow B_2$ 
is described \FMW\ 
by assuming the ${\bf P}^1$ bundle over $B_2$ (with section $r$) 
to be a projectivization of a vector bundle $Y={\cal O}\oplus {\cal T}$, 
where ${\cal T}$ is a line bundle over $B_2$ and the cohomology class 
$t=c_1({\cal T})$ encodes the ${\bf P^1}$ fibration structure.
Let $N$ denote the normal bundle of $W$ in $B_3$ with $c_1(N)=-t$ from 
$rr=-tr$ (on the heterotic side $t$ indicates an asymmetry
between the cohomological data of the two bundles \FMW\ ). 
To explain this relation and also for later use let us look into
the relevant geometry. Let ${\cal O}(1)$ be the line
bundle on the total space of ${\bf P}(Y)\rightarrow B_2$ which is fibrewise
the usual ${\cal O}(1)$.  
Let $a,b$ be homogeneous coordinates of ${\bf P}(Y)$ and think of 
$a,b$ as sections, respectively, of ${\cal O}(1)$ and 
${\cal O}(1)\otimes{\cal T}$
over $B_2$ with Chern-classes $r=c_1({\cal O}(1))$ and
$r+t=c_1({\cal O}(1)\otimes {\cal T})$. Then the cohomology ring of $B_3$ is 
generated over the cohomology ring of $B_2$ by $r$ with the relation 
$r(r+t)=0$ (meaning that the divisors $a,b$ which are dual to $r$ resp. $r+t$
do not intersect). The Chern-classes of $B_3$ are then computed by applying 
the adjunction formula
$c(B_3)=(1+c_1+c_2)(1+r)(1+r+t)$ (here unspecified Chern classes refer to 
$B_2$: $c_i=c_i(B_2)=\pi^*(c_i(B_2))$).}
over $B_2$ in such a way that $X$ is $K3$ fibered over $B_2$.
Both of these models
require in general for consistency the inclusion of some brane-impurities: 
the heterotic string leads to some five-branes because of anomaly cancellation
\FMW\ , and the $F$-theory vacuum leads to a 
number of space-time filling three-branes 
because of tadpole cancellation \SVW. These three-branes are located at
points on the base $B_3$. 

Let us briefly recall the possible contributions
to the tadpole equation. The first one comes from the coupling \dlm\ 
\eqn\anompol{-\int_{{\bf R}^3\x X} C\wedge I_8}
Here we write the term on the level of M-theory.
Corresponding terms  involving a coupling of $B^{NS}$  with an 8-form
in curvature in type IIA give rise to a 2D term \vw. Lifting of such a
term to F-theory is also possible \fms, and  gives rise to the
above-mentioned tadpole. Indeed using $ \int_X I_8 = \chi(X)/24$ \bb,
one finds for the number of three-branes 
\eqn\euler{n_3=\chi(X)/24}
Similarly when one takes into account non-vanishing four-fluxes, the
Chern-Simons coupling 
\eqn\CS{\int_{{\bf R}^3\x X} C\wedge G\wedge G}
also contributes to the number of three-branes \damu\ ($G={1\over 2\pi}dC$)
\eqn\mitflux{{\chi(X)\over 24}=n_3+{1\over 2}\int_X G\wedge G}
Here we consider the $F$-theory analog of $M$-theory four-flux in the
limit that the area of the elliptic fibers is very small. The 
four form $G$ on $X$ can be expanded in terms of forms
denoted by $p_2, H_3$ and $g_4$ (subscripts indicating the degree) \GVW
\eqn\gvs{G=p_2\wedge \chi+\sum_i H_3^i\wedge \theta^i\;\; +g_4}
where $\chi$ is an integral two-form generating the two-dimensional 
cohomology of the fibers and $\theta^i$, $i=1,2$, is a basis of
$H^1(T^2)$. Setting $g=p=0$ guarantees 
that $G$ is part of the primitive $H^{2,2}(X)$ cohomology and is odd under the
fiber involution. In type IIB theory $H$ can be expressed
in terms of the NS and Ramond three-form field strength $H^{NS}, H^R$ and 
chosen to be an integral $(2,1)$ form, 
well defined up to $Sl_2({\bf Z})$ transformations around the 
seven-brane loci. So $G\wedge G$ goes in the $F$-theory/type IIB limit to
\foot{We will not discuss here the special case where 
the four-flux is localized 
around fibers over $W$ and it lifts  to the field strength 
of the corresponding gauge field (cf. \drs\ ) with
$G\wedge G$ leading to the term $c_2(E)$.}
\eqn\flux{{1\over 2}\int_X G\wedge G\rightarrow 
\int_{B_3}{1\over {\tau_2}} H\wedge {\bar H}}
with $H=H^R-\tau H^{NS}$ and ${\bar H}=H^R-{\bar\tau}H^{NS}$.

In  case the point $p$ (which constitutes 
the compact part of the world-volume of a three-brane) lies not only
in $B_3$ but actually in $W$, one can consider $p$ as a 
small instanton of an unbroken gauge group located in the (compact part $W$ 
of the) corresponding seven-brane world-volume, which corresponds to a
component of the discriminant locus of the 
elliptic fibration. The discriminant locus in general decomposes into
components (resp. seven-branes); for example matching the perturbative
heterotic gauge content requires having three components, two of  which
carry the  (unbroken heterotic) gauge group $G_1\times G_2$, and the third
one - $I_1$ singular fibers. Here we will think of a situation where 
we have only two seven-branes with $G_1$ over $W$ and $I_1$ singularity 
located over $D_1$, corresponding to a heterotic bundle $(V_1,E_8)$) 
\AC.

Note that in general the seven-branes are intersecting. In our
context this means that the component $W$ of 
the discriminant which carries the $G_1$ singularity will intersect the 
component carrying the $I_1$ singularity. However, since only on 
$W$ a Chan-Paton bundle $E$ is switched on (the group of the $I_1$
surface is $U(1)$) at the moment we do not have to worry about
intersections (but see remark 2).

One can also have an instanton in a rank $rk(E)$ bundle $E$
over $W$, breaking part of the $F$-theory gauge group $G$ associated with
$W$ \bersh. Then the possibility of transitions where such an instanton
becomes point-like and leaves the (part of the) discriminant surface
$W$ in $B_3$ arises, leading to a further modification of the consistency
condition
\bersh\
\eqn\micha{{\chi(X)\over 24}=n_3+\int_W c_2(E)+
\int_{B_3}{1\over {\tau_2}} H\wedge {\bar H}}
The last possibility fits with the general philosophy that one always has
in connection with a D-brane a vector bundle over its world-volume.
Moreover, one  can nicely describe transitions associated with small
instantons using this formula, balancing the instanton number with the
number of three-branes.

However, it is known that the D-brane charge is
given in terms of the topology of the embedded worldvolume $W$ ($i: W
\hookrightarrow B = B_3$ )  \mm\ and the topology of $E$
\eqn\kcharge{q = {\rm ch} (i_!E) \sqrt{\hat A(B)}.}
and thus in general one has to take into account in \micha\ 
the non-trivial effects associated with the embedding $i$. 
Then a  general contribution
to the tadpole is no longer just given by integrating, as in 
$\int_W c_2(E)$, the top class of the
bundle over the compact part of the world-volume, but should be deduced
from  the D3-brane coupling to D7  and is of the form
\eqn\taddel{- \int_{\bf R^4}A_4 \int_W Y_4}
Here $Y_4$ denotes the degree four part of the anomalous coupling $Y$
given by \refs{\ghm, \chy, \mm, \szabo} 
\eqn\greg{Y = ch(E)e^{-{1\over 2}c_1(N)}\sqrt{{\hat{A}(W)}\over
{\hat{A}(N)}}}
(with the  A-roof genus given by $\hat{A}=1-{p_1\over 24}+\cdots $ and the
Chern character $ch=rk+c_1+{c_1^2\over 2}-c_2+{c_1^3-3c_1c_2+3c_3\over 6}
+\cdots $).
Note that one derives \kcharge\ by application of Grothendieck-Riemann-Roch
theorem to the anomalous couplings of RR fields to $Y$ on D-brane
worldvolume. 

In summary, because of the anomalous 
D-brane charge, the contribution $\int_W c_2(E)$ has to be replaced by
\eqn\anomcharge{-\int_W \Bigg( rk(E)\Big( {c_1(N)^2\over 8}-{p_1(W)\over 48}+
{p_1(N)\over48}\Big) -{c_1(E)c_1(N)\over 2}+ch_2(E)\Bigg)}
Since we are concerned here only with sevenbranes, 
$p_1(N)=c_1(N)^2$ for the line bundle $N$. 
With these corrections \micha\ becomes 
\eqn\newtotal{{\chi(X)\over 24}=n_3-\int_W \Biggl(ch_2(E)+
rk(E)\left({c_1(N)^2\over 8}+{c_1(N)^2\over 48}
-{\sigma (W)\over 16}\right)
- {c_1(E)c_1(N)\over 2}
\Biggr)
+\int_{B_3}{1\over {\tau_2}} H\wedge {\bar H}}
As noted already there are different groups of (intersecting) sevenbranes,
and one has to sum over the ``instanton" contributions from all of them
(including those with trivial gauge bundles), so a sum is assumed in the
second term on the right hand side.

In order to find an interpretation of \newtotal, we once more use the
Grothendieck-Riemann-Roch theorem (see \Karo) for a holomorphic map
$i:W\rightarrow B_3$, which in our notation reads 
(with $Td = {\hat A} e^{{1\over 2}c_1}$)
\eqn\grr{i_*(ch(E)Td(W))=ch(i_!E)Td(B_3).}
where $i_!$ is the $K$-theoretic Gysin map 
(a homomorphism $K(W) \rightarrow K(B_3)$) and can be defined as 
\eqn\gys{i_!E:=\sum_q(-1)^qi^q_*E}
where $i^q_*E$ are direct image sheaves (the $q$-th direct image of $E$).
Since $i:W\rightarrow B_3$ is an embedding we have $i^q_*E=0$ for $q>0$
and $H^p(W, E)\cong H^p(B_3, i^0_*E)$ for $p\ge0$ (cf. \Karo). Then \gys\ 
simplifies for our case to $i_!E = i_*E$  and we have 
to deal only with the torsion sheaf $i_*E$.

For further application to \newtotal, we use
Gysin homomorphism $i^c_*$ for cohomology which maps classes of
codimension
$p$ in $B_2 = W$ into classes of codimension $p$ in $B_3$. Denoting the
Poincar\'e duality on $B_3$ by $D_{B_3}:H^p(B_3)\rightarrow H_{6-p}(B_3)$
(and similarly on $W$, $D_{W}:H^p(W)\rightarrow H_{4-p}(W)$), we can
define
the action of $i_*$ on $H^*(W, {\bf Q})$ as $i^c_*=D_{B_3}^{-1}i^h_*D_{W}$
where $i^h_*$ is the usual map induced on homology. It is easy to see that
in our case $i^c_*:H^p(W, {\bf Q}) \rightarrow  H^{p+2}(B_3, {\bf Q})$
(for
example $i^c_* 1=r$ where $r$ as above denotes the class of $W$ in $B_3$).
If $i: W \hookrightarrow B_3$ is an embedding of $W$ as a submanifold of
$B_3$, $Td(W) = (Td(N))^{-1}  i^*Td(B_3)$. We can use now that for $\phi
\in H^*(W, {\bf Q})$ and $\eta \in H^*(B_3, {\bf Q}),$ $i^c_*(\phi \wedge
i^* \eta) = i^c_*\phi \wedge \eta$ and see that \grr\ implies the
Riemann-Roch theorem for an embedding 
\eqn\rre{ch(i_!E) = i^c_*(ch(E) (Td(N))^{-1})
.} 
expansion of \rre\ gives for the Chern characters of the torsion sheaf
$i_*E$ 
\eqn\chc{\eqalign{ ch_1(i_*E)&= rk(E)r\cr 
ch_2(i_*E)&= i^c_* \Big(c_1(E)- rk(E){c_1(N)\over 2}\Big) \cr 
ch_3(i_*E)&= i^c_*\Big(ch_2(E)+rk(E)({c_1(N)^2\over 8}+{c_1(N)^2\over 24})
-
c_1(E){c_1(N)\over 2}\Big) }} 
Comparison to \newtotal\ shows that the changes in the tadpole equation
can be compactly written as 
\eqn\change{\int_W c_2(E) \longrightarrow -\int_B ch_3(i_*E) + {1\over
48}rk(E)
\int_B r\wedge p_1(B)}
and the  general condition for
the tadpole cancellation in the case of nontrivial D-brane embeddings as 
(once more with a sum over all brane contributions is assumed)
\eqn\newt{{\chi(X)\over 24}=n_3 - \int_{B_3}ch(i_*E) \sqrt{{\hat
A}(B)} + \int_{B_3}{1\over
{\tau_2}} H\wedge {\bar H}.} 
Of course one could have arrived at \newt\ directly from \greg, 
however we have chosen a lengthier
presentation shown above since we will need some of
the explicit formulae such as \newtotal\ and \change\ for future use.
We see that the original instanton charge
$c_2(E)$ \bersh\ is replaced by $\left[ch(i_*E) \sqrt{{\hat
A}(B)} \right]_3$ (or as explained in \mm, $i_*E \in K(B)$).
Note that in the spirit of the
$F$-theory (or better type IIB) reinterpretation of the former $M$-theory
relation and also in the spirit of $K$-theory interpretation of D-branes 
\refs{\mm, \wik}, the right hand side is now expressed completely
on terms of the base $B_3$ visible to type IIB (and not in terms of $X^4$ 
or $W$); the left hand side which
seems still to involve the complete four-fold not visible to type IIB can
also be expressed in terms of data visible to type IIB using the
stratification of singularities \AC, \SVW\ .

Finally let us make some remarks on the modification \newt\ of the 
tadpole equation concerning the question of transitions, some
integrality conditions and the moduli space of $i_*E$ (which
is also relevant for the duality with the heterotic string). To be somewhat
concrete we turn to the well studied example mentioned above, the
standard case of Hirzebruch surfaces  $B_2={\bf F}_n$ where some
simplifications occur as the first
Pontrjagin class $p_1(W)=c_1(W)^2-2c_2(W)=3\sigma(W)$ vanishes.
Although it is not important for the conceptual understanding, one may
as a minor technical simplification also make the standard assumption 
that for the gauge bundle $c_1(E)$ vanishes.\foot{Note however that this is
possible only on the ${\bf F}_n$ with even $n$ since when $W$ is not a
spin manifold $c_1(E)$ has to be odd in order to define a $Spin^c$ structure
\fw\ }.  With all this taken into account, we rewrite once more the change in
the tadpole condition for the topological charge
of the Yang Mills instanton  (including 'gravitational' contributions)
\eqn\correction{c_2(E)\;\; \longrightarrow \;\;
c_2(E)-rk(E)({c_1(N)^2\over 8}+
{c_1(N)^2\over48})}
\medskip

\noindent
{\it 1. Transitions $\,$} Note first that this change doesn't
constitute any problem with
respect to the seemingly already perfectly 'balanced' terms
$n_3+c_2(E)$ in the original equation. The reason is simply that the
correction term is only an 'once and for all' background term,
i.e. the tadpole equation has to be balanced with this term included and
then it will not change in transitions as strictly speaking $rk(E)$
will not change (but of course $E$ will become reducible when the
instanton number $c_2(E)$ is lowered in transitions producing points
contributing to $n_3$). So we work in a set-up where the rank
$rk(E)$ of the bundle,
which is possibly turned on, is always the maximal one, i.e. $rk(E)=rk(G)$ for
the gauge group $G$ which comes from the $F$-theory data and which is
eventually partially broken by the embedded bundle $E$. Note that $rk(G)$
is the multiplicity $k_W$ of the surface $W$ in the discriminant divisor,
resp. the number of seven-branes which have coalesced.

\medskip 

\noindent
{\it 2. Integrality properties $\,$}  There are some subtle integrality
issues pertaining to this
formula. The charge \kcharge\ and thus the modification of the tadpole
condition are not in general in integral cohomology, and thus one may
wonder about the
integrality properties and the consistency of the tadpole condition.
Indeed there are two correction terms in \correction, whose integrality
properties we would like to understand: the first one 
$rk(E){c_1(N)^2\over8}$ comes just from the term 
$e^{-c_1(N)/2}$ in \greg; the second
one $rk(E){c_1(N)^2\over 48}$ comes from the term
$1/\sqrt{\hat{A}(N)}$ in \greg\ . 

We start by recalling  that the LHS of the tadpole equation \newt\ is in 
${1\over 4} {\bf Z}$ as the Euler number of $X^4$ is divisible by $6$
\SVW.  On the other hand the RHS is - as far as the flux term is concerned
-  in ${1\over 4} {\bf Z}$ too because (although $G$ can be in the 
half-integral cohomology as its quantization law \wflux\ demands that
$G-{c_2(X^4)\over 2}$ lies in the integral cohomology)
one finds\foot{Note that 
the argument cannot use evenness of the
intersection form as this is given only for $c_2(X^4)$ even, which
according to the quantization law is exactly not the critical case; 
instead one argues from $G\equiv c_2(X^4)/2$ (mod
$H^4(X^4,{\bf Z})$) that
${1\over 2}\int_X G\wedge G\equiv {1\over 8} \int_Xc_2^2(X^4)$
(mod ${1\over 2}{\bf Z}$); but 
${1\over 8} \int_Xc_2^2(X^4)\in {1\over 4} {\bf Z}$ as 
$\int_Xc_2^2(X^4)=480 +{\chi (X^4)\over 3}\in 2{\bf Z}$ (cf. \SVW\ ).} 
that 
${1\over 2}\int_X G\wedge G$, which would seem to lie in ${1\over 8} {\bf Z}$, 
also lies in ${1\over 4} {\bf Z}$. 
Actually the proof of this fact (cf. the last footnote) shows more, 
namely that even in the general case of
non-integral ${\chi (X^4)\over 24}$ nevertheless an half-integral number 
for the three-brane/bundle contributions 
is predicted by having $G$ included: for 
${\chi (X^4)\over 24}-{1\over 2}\int_X G\wedge G$ is congruent to an
integer mod ${1\over 2}{\bf Z}$, i.e. is itself in ${1\over 2}{\bf Z}$:
\eqn\integr{{\chi (X^4)\over 24}-{1\over 2}\int_X G\wedge G
\in {1\over 2}{\bf Z}}

How is this occurence of half-integrality to be explained ? Clearly
$n_3$ and $c_2(E)$ should be integral. However we will see that 
the gravitational contributions (which are related to the geometry of
the (compact part $W$ of the) 
seven-brane resp. its embedding) come out only half-integral. More
precisely we will see that the first correction term accounts for 
the non-integrality (but is still half-integral) whereas the second one is
actually integral (at least in the case under consideration $W={\bf F}_n$,
and is half-integral in general).

So let us after this prediction \integr\ 
from the tadpole equation investigate whether the remaining
terms, especially the new ones satisfy this integrality requirement. 
The base $B_3$ of the type IIB theory is assumed to be 
a spin manifold. We find from $c_1(B_3)\equiv 0$
$(mod \ 2)$ and from  $c_1(B_3)=c_1+2r+t,$ computed from the adjunction 
formula (see footnote 4), that $t\equiv c_1$ $(mod \ 2)$ and so
$t^2\equiv c_1^2$ $(mod \ 4)$; since
$c_1^2=8$ for Hirzebruch surfaces\foot{For more general $B_2$ as del
Pezzo surfaces $dP_k$ (k-fold blow-up of $P^2$) or
Enriques surface ${\cal E}$ of $c_1^2(dP_k)=9-k$ resp. $c_1^2({\cal E})=0$ 
the condition $c_1^2 \equiv 0 (mod 4)$ shows that
only the bases  $dP_1=F_1, dP_5$, $dP_9$ and Enriques are
unproblematical.} 
one has $t^2\equiv 0$ $(mod \ 4)$ and so the first correction term 
$rk(E){t^2\over 8}$
is in ${1\over 2} {\bf Z}$ (even in the case of $c_1(E)\neq 0$).
So the first correction term 'explains' the non-integrality but 
does not lead to further problems as it is half-integral (but see the last
footnote).

On the other hand the second correction term clearly seems to destroy the
demanded (half-)integrality properties in general. To understand this, 
we have to turn now to the global consistency conditions. So far 
we have merely focused  on the deviation from the simple
result $c_2(E)$ for the bundle/instanton contribution for one given
bundle. 
However as already remarked there are necessarily groups of non-parallel
seven-branes, and thus the contributions from all of them have to be taken
into account. In particular, in the simplest example one has in $B_3$ at
least the surface $D_1$ of $I_1$ singularities as further
component of the  discriminant surface. We recall that since the
total space $X_4$ is a
Calabi-Yau manifold, there is a global restriction on the seven-branes,
namely the Kodaira condition
${1\over 12}\sum_{W_A}k_Ar_{W_A}=c_1(B_3)$
(where we sum over all discriminant components $W_A$ of cohomology
class $r_A$ and multiplicity $k_A$) (cf. \MV\ ). 
Using \change\ and\foot{Strictly speaking we argued for this only for D-branes;
however for more general $(p,q)$ seven-branes the couplings relevant for our
investigation should be still the same by $Sl(2,{\bf Z})$ duality. By
virtue of this duality we impose the 'maximal rank condition'
$rk(E_A)=k_A$ also for the other seven-branes (even if these do not
contribute to the resulting four-dimensional gauge fields).} 
$rk(E_A)=k_A$, this gives exactly the condition
necessary for understanding the needed integrality properties:
\eqn\kconsist{p_1(B_3){1\over 48}\sum_{W_A}rk(E_A)r_{W_A}
={1\over 4}p_1(B_3)c_1(B_3)}
(which could be even further evaluated as 
${1\over 2}p_1(B_3)r={1\over 2}(p_1(B_2)+t^2)$). Since $B_3$
is a spin manifold, i.e. of even $c_1$, its $p_1=c_1^2-2c_2$ is even too,
leading to two factors of $2$. 
Thereby this contribution is actually integral.\foot{Since we have used 
$c_1^2(N)=
p_1(B_3)|_W$, this argument may seem
somewhat special for  the case of ${\bf F}_n$. In general, for
$W_A\ne B_2={\bf F}_n$, and thus
$p_1(W) \ne 0$,
one may still worry about the integrality properties of $ch_3(i_*E)$. 
However by yet another rewriting using the adjunction formula one can
isolate a part  $\sum_{W_A}rk(E_A)r_{W_A}\wedge p_1(B)/24$ with
the remaining part being a sum
of an (integral) index
$\int_W {\hat A(W)}e^{c_1(W)/2} ch(E)$ and
 half-integral terms (using $c_1(B)$ even for $B$ spin).}

\medskip

\noindent
{ \it 3.  The moduli space $\,$} As we have
exchanged the bundle
$E$ by the object $i_*E$
when using  the 'three-dimensional' expression $\left[ch(i_*E)
\sqrt{{\hat A}(B)} \right]_3$ 
instead of the instanton contribution let us finally consider the
influence 
on the moduli space of this shift from $E$ to $i_*E$. 
We will find that the dimension of the 
associated moduli space is unchanged as required by the duality with 
heterotic string \bersh. Whereas in \bersh\ the dimension of the 
moduli space of $E$ over $W$ was 
computed 'intrinsically' (only with respect to $W$) and was given by the 
dimension of $H^1(W,End(E)),$ we have to consider now the torsion sheaf
$i_*E$ which lives on $B_3$. The dimension of the associated moduli
space is given by the dimension of $Ext^1_{B_3}(i_*E,i_*E)$. One can in 
general expect that the dimension of the moduli space associated to $i_*E$ 
is bigger than the dimension of the moduli space of $E$ over $W$. This can
happen because, naively speaking, $W$ can move inside $B_3$ and we therefore 
have additional deformations. The number of deformations of  $W$ in
$B_3$ is simply given by the number of sections of the normal bundle, i.e. the 
dimension of $H^0(N)$. This naive picture can be made precise by considering 
the long exact sequence (first written down and proven in \thom
)\foot{We thank R. Thomas for helpful discussions on this.}
\eqn\seq{0\rightarrow Ext^1_W(E,E)
          \rightarrow Ext^1_{B_3}(i_*E,i_*E)\rightarrow Ext^0_W(E, E\otimes N)
          \rightarrow Ext^2_W(E,E)\rightarrow}
Since $E$ is assumed to be a vector bundle over $W$ we have
isomorphisms $Ext^1_W(E,E)= H^1(W,End(E)), Ext^0_W(E, E\otimes N)= 
Hom(E, E\otimes N)$ and $Ext^2_W(E,E)= H^2(W,End(E))$ leading to
\eqn\seq{0\rightarrow H^1(W,End(E))
          \rightarrow Ext^1_{B_3}(i_*E,i_*E)\rightarrow Hom(E, E\otimes N)
          \rightarrow H^2(W,End(E))\rightarrow}
using the fact that $H^i(W,End(i_*E))=H^i(W,i_*End(E))$ 
and that $H^i(W,i_*End(E))=H^i(W,End(E))$ \hart.

One assumes in general (cf. \bersh\ ) 
that $E$ is a {\it good} instanton bundle
so that $H^2(End(E))=0$ which can be thought as a condition to get a 
smooth moduli space whose dimension can be evaluated by \koba. One can follow
the conditions under which $H^2(W,End(E))$ 
vanishes; for this we decompose $H^2(End(E))$ into its trace-free 
part $H^2(End(E)_0)$ and in $h^{(0,2)}(B_2)$. Since we 
consider rational $B_2$'s
we have $h^{(0,2)}(B_2)=0$ so we are left with the trace-free part. If we
assume a high enough instanton number of $E,$ a theorem of 
Donaldson \donal\ states that
this term vanishes too and we can consider the exact sequence
\eqn\seq{0\rightarrow H^1(W,End(E))
          \rightarrow Ext^1_{B_3}(i_*E,i_*E)\rightarrow Hom(E,E\otimes N)
          \rightarrow 0}
which gives 
\eqn\dim{{\rm dim} \ Ext^1_{B_3}(i_*E,i_*E)={\rm dim} \ H^1
(W,End(E))+{\rm dim} \ H^0(Hom(E,E\otimes N))}
Now one has $H^0(Hom(E,E\otimes N))=H^0(Hom(E,E))\otimes H^0(N)$ but for  
our normal bundle of $c_1(N)=-t$ we have \foot{As
(assuming $t$ effective) any
non-trivial element of $H^0(N)$ would give, multiplied by some 
non-constant section (having zeroes) of $N^{-1}$ (which exists as
$c_1(N^{-1})=t$), a non-constant element of $H^0({\bf C})={\bf C}$.}
$Hom(E,E\otimes N)=0$.
So we finally find that the dimensions of the moduli spaces 
of $i_*E$ and $E$  match. 

\bigskip
\bigskip
\centerline{\bf Acknowledgements}

We would like to thank A. Klemm, G. Moore, S. Theisen, R. Thomas, C. Vafa
and S.-T. Yau for discussions. BA thanks the Harvard University and RM
thanks the Erwin Schr\"odinger Institute for hospitality during part of
this work. 

\listrefs
\bye